\documentstyle[version2,preprint,aps]{revtex}
\begin{document}
\draft
\begin{title}
Total Cross Sections for Neutron Scattering
\end{title}
\author{ C.R.~Chinn$^{(a),(b)}$, Ch.~Elster$^{(c)}$,
 R.M.~Thaler$^{(a),(d)}$, and S.P.~Weppner$^{(c)}$.}
\begin{instit}
$^{(a)}$ Department of Physics and Astronomy, Vanderbilt University,
Nashville, TN  37235
\end{instit}
\begin{instit}
$^{(b)}$ Center for Computationally Intensive Physics,
Oak Ridge National Laboratory, \\ Oak Ridge, TN  37831
\end{instit}
\begin{instit}
$^{(c)}$ Institute of Nuclear and Particle Physics,  and
Department of Physics, \\ Ohio University, Athens, OH 45701
\end{instit}
\begin{instit}
$^{(d)}$ Physics Department, Case Western Reserve University, Cleveland,
OH  44106.
\end{instit}
\vspace{10mm}
\receipt{\today}
\vspace{1in}
\pacs{PACS: 25.40.Cm}
\pagebreak
\begin{center}
ABSTRACT
\end{center}
\begin{abstract}
Measurements of neutron total cross-sections are both extensive and
extremely accurate. Although they place a strong constraint on
theoretically constructed models, there are relatively few
comparisons of predictions with experiment.  The
total cross-sections for neutron scattering from $^{16}$O and
$^{40}$Ca are calculated as a function of energy from
$50 - 700$~MeV laboratory energy
with a microscopic first order optical potential
derived within the framework of the Watson expansion. Although
these results are already in qualitative agreement with the data,
the inclusion of medium corrections to the propagator
is essential to correctly predict the energy dependence given by
the experiment. In the region between $100$ and $200$~MeV, where
off-shell t$\rho$ calculations for both $^{16}$O and $^{40}$Ca
overpredict the experiment, the modification
due to the nuclear medium reduces the calculated values.
Above ~300 MeV these corrections are very small and depending
on the employed nuclear mean field
tend to compensate for the underprediction of the off-shell
t$\rho$ results.
\end{abstract}
\pagebreak
%****************************************************************************
\narrowtext
%******************************************************************
\hspace*{10mm}
Accurate measurements of total cross-sections for the scattering of
neutrons from a variety of nuclei provide stringent constraints on
theoretical models.
In this sense one would like to incorporate a realistic
physical description of the nucleon-nucleon [NN] interaction,
which quantitatively reproduces the NN observables,
including
the total neutron cross-section, into a theoretical model of
neutron-nucleus scattering.  Such a description would ideally
result from multiple scattering theory and the use the NN interaction
together with nuclear wave functions as input to obtain nucleon-nucleus
observables.
In a recent publication \cite{medium}
we presented a microscopic treatment of the modification of
the NN propagator due to the nuclear medium,
formulated to be consistent with the first order spectator
expansion \cite{PT,Sicil}. Calculations of proton and neutron
elastic scattering from $^{40}$Ca in the first order spectator
expansion were presented, in which the effect of the nuclear
medium was taken into account. These results were very encouraging.
In particular, the neutron total cross-section for scattering
from $^{40}$Ca was shown as a function of energy.  It was
gratifying to observe \cite{medium} that the effect of the
nuclear medium, although not large, lowered (relative to those for
which a free NN t-matrix was employed) the calculated values
in the energy regime between $100$ and $300$~MeV
and brought the theoretical
predictions into closer agreement with the measured values.
Furthermore, above $\sim$400 MeV, where the `free' predictions fell below
the experimental data, the medium correction raised the predicted
values, so that in this region a better description of the
experiment was also achieved.
The comparisons shown in Ref. \cite{medium} were limited to the
scattering of neutrons from $^{40}$Ca, and it was felt that
it would be reassuring to extend the calculation to other targets.
At present the calculations are limited to spin-saturated
even-even targets and so only the scattering from $^{16}$O and
$^{40}$Ca are investigated here.

\hspace*{10mm}
In this work the full Bonn potential \cite{Bonn} is used to
calculate the ${\rm H}(n,p)$ total cross-sections for
projectile energies less than 350~MeV.  This interaction includes
the effects of relativistic kinematics, retarded meson propagators
as given by time-ordered perturbation theory, and crossed and
iterative meson exchanges with $NN$, $N\Delta$ and $\Delta\Delta$
intermediate states.  For energies greater than 350~MeV a less
sophisticated
high energy extension of the Bonn potential \cite{D52} is used,
which incorporates the effects of pion production via $N\Delta$ and
$\Delta\Delta$ iterative diagrams.

\hspace*{10mm}
The calculations for nuclei other than $^1$H are performed in
the parameter-free manner as outlined in Ref. \cite{medium},
where the only inputs are the free NN interaction, target
nuclear densities and the static mean field potential used to
model the medium effects. The free NN t-matrix is taken from the
same Bonn potential that is used to calculate the ${\rm H}(n,p)$ result.
The nuclear mean field potentials are taken from a
Hartree-Fock-Bogolyubov (HFB) microscopic nuclear structure calculation,
which utilizes the density-dependent finite-ranged {\it Gogny D1S}
effective NN interaction \cite{HFB,Gogny}.
A second choice involves a nonrelativistic reduction of the mean
field potentials resulting from a Dirac-Hartree (DH) calculation based upon
the $\sigma$-$\omega$ model \cite{DH}.
The neutron-nucleus calculations are performed in an `off-shell t$\rho$'
framework using the optimum factorized form as described in
Ref.~\cite{Wolfe}, where the fully off-shell effective NN
t-matrix is used, but with diagonal nuclear densities.

\hspace*{10mm}
For a comparison with two-nucleon scattering, in Fig.~1 the
total cross-section for neutron-proton $(np)$ scattering is
shown. The solid line represents the experimental
data \cite{Ullman}, while the diamonds represent the predictions
from the full Bonn interaction \cite{Bonn}. It should be noted
that, although these data were {\underline {not}} included in the fitting
procedure for the Bonn potential, excellent agreement is
obtained.  Included in Fig.~1 are the predictions for
the neutron-neutron $(nn)$ total cross-section, which are obtained
by considering only the T=1 partial waves. No data are given for
this case, since none exist.  However, the excellent agreement with
the $(n,p)$ experimental data suggests that the $(n,n)$ predictions
are likewise reliable.

\hspace*{10mm}
In Fig.~2 the total neutron cross-section data for scattering
from  $^{16}$O and $^{40}$Ca are presented as a function of
the neutron laboratory energy.  Note that for $^{16}$O in the
upper panel of Fig.~2
the corrections due to the nuclear medium are small, but tend
in the correct direction. At lower energies ($100-200$~MeV),
these medium effects reduce the calculated cross-section in
comparison to the free result and bring the predictions into rather
good agreement with the measured values. At higher energies the effect
of the coupling of the struck nucleon to the rest of the nucleus tends
to increase the total cross-section value slightly to again yield a result
which is closer to the data. The analogous result for  $^{40}$Ca is
shown in the lower panel of Fig.~2, and as described for
$^{16}$O, the contributions due the
propagator modification by the nuclear medium tend to improve the
description of the experimental observable.

\hspace*{10mm}
If one were to assume that there were no shadowing effect
then one would expect the total neutron cross-sections to
scale with the number of nucleons, $A$.  In this case the
isospin averaged $\langle\sigma_T\rangle$ of the $(n,p)$
and $(n,n)$ cross-sections shown in Fig.~1 multiplied by
$A=16$ and by $A=40$ are indicated by asterisks in the
upper and lower panels of Fig.~2, respectively.  These points
would represent the energy dependent total cross-sections in the
absence of shadowing and thus correspond to an upper limit on the
possible values of the total cross-sections.
However, the measured values for $^{16}$O and $^{40}$Ca are indeed of
the order of $25\%-30\%$ smaller than this limiting value, and one
should expect any microscopic treatment to be closer than this limit.
The horizontal arrows indicate the geometric black disk cross-section
$2\pi R^2$ (with $R=\frac{5}{3}\langle r_{rms}\rangle$).
For energies for which the total
cross-section exceeds this geometric limit, one should be very
skeptical of the first-order multiple scattering theory.  For
$^{16}$O this energy is roughly $50$~MeV and for $^{40}$Ca it is
about $70$~MeV.

\hspace*{10mm}
Although the $^{16}$O and $^{40}$Ca results do not scale as
$A\times\langle\sigma_T\rangle$,
they may scale for some effective number of nucleons, $A_{eff}$.
In this sense one can partially account for shadowing.  In
Fig.~3 the average neutron-nucleon total cross-section,
$\langle\sigma_T\rangle$ is plotted along with the the $^{16}$O
and $^{40}$Ca results divided by $A_{eff}$.  In this case
$A_{eff}\approx 12$ and $A_{eff}\approx 26$ for $^{16}$O and $^{40}$Ca,
respectively.  The $^{16}$O and $^{40}$Ca experimental
measurements scale remarkably well for energies greater than
about $80$~MeV.  For the high energies where scaling is
occurring it seems clear one is basically observing effects due
to the NN interaction.  Contributions beyond the first order multiple
scattering theory are not so important in this regime.  It is also
true that this may be an indication that the total cross-section
data is less revealing than angular measurements.  For smaller
energies the break down of scaling illustrates how higher order
corrections to the first order theory are required.

\hspace*{10mm}
In Fig.~4 the total elastic cross-sections $\sigma_{el}$ and the total
reaction cross-sections $\sigma_{R}$ for $^{16}$O are displayed
separately. Since the reaction cross-section is considerably larger
than the elastic cross-section for laboratory kinetic energies greater
than $100$~MeV, it follows that the corrections to the total cross-section
due to the nuclear medium  are largely given by contributions to the
reaction cross-sections.  The same conclusion can be made about
the $^{40}$Ca elastic and reaction cross-sections.  Qualitatively
the observed effect of the medium may be interpreted as follows.
Since the interactions
between the struck nucleon and the `residual (A-1) nucleus' is
attractive, the medium correction in a sense tends to reduce the
knock-out  probability for outer nucleons, thus reducing the reaction
cross-section.
This is particularly the case for energies lower than 200 MeV, where the
scattering process is expected to be surface dominated.
At higher energies , the projectile neutron penetrates
more deeply into the nucleus, so that this effect becomes less important.

\hspace*{10mm}
To demonstrate that contributions due to the nuclear medium can be
significant in neutron elastic scattering, the neutron differential
cross-section, the analyzing power and the spin rotation parameter for
scattering from $^{16}$O at $100$ and $500$~MeV are shown in
Figs.~5 and 6.  At higher energies the
corrections do not begin to manifest themselves until the scattering
angles become so large that higher order processes need to
be taken into account. At the lower energy there is certainly a
significant difference in the spin observables even at small angles.

\hspace*{10mm}
In conclusion the precision neutron total cross-section measurements
provide another striking confirmation that the first order
theory of the optical potential can accurately describe data in the
appropriate regime of applicability.  These successes encourage us to
proceed in this direction and in the future to include higher order
contributions in the multiple scattering spectator expansion.

\vspace{5mm}

\vfill
\acknowledgments
The authors would like to thank R.W.~Finlay and J.~Rapaport
for many discussions concerning this work. The computational
support of the the Ohio Supercomputer Center under
Grants No.~PHS206 and PDS150 is gratefully acknowledged.
This work was performed in part under the auspices of the U.~S.
Department
of Energy under contracts No. DE-FG02-93ER40756 with Ohio University,
DE-AC05-84OR21400 with Martin Marietta Energy Systems, Inc., and
DE-FG05-87ER40376 with Vanderbilt University.  This research has also
been supported
in part by the U.S. Department of Energy, Office of Scientific Computing
under the High Performance Computing and Communications Program (HPCC)
as a Grand Challenge titled the Quantum Structure of Matter.

%----------------------------------------------------

\pagebreak

%%%%%%%%%%%%%%%%%%%%%%%%%%%%%%%%%%%%%%%%%%%%%%%%%%%%%%

\pagebreak
%%%%%%%%%%%%%%%%%%%%%%%%%%%%%%%%%%%%%%%%%%%%%%%%%%%%%%
\noindent
\figure{ The total cross-section $\sigma_{tot}$ for the scattering of
         neutrons from protons ${\rm H}(n,p)$ is shown.  The solid curve
         corresponds to the experimental data \cite{Ullman}.
         No error bars are shown, since the total systematic uncertainty
         is smaller than 1\% and the errors fall within the thickness of
         the plotted line.
         The diamonds represent the theoretical predictions from the
         full Bonn potential,
         the triangles give the results for the
         $(n,n)$ case.  \label{fig1}}

\figure{ The total neutron-nucleus total cross-sections for
         scattering from $^{16}$O and $^{40}$Ca are shown as
         a function of the incident neutron kinetic energy in the
         upper and lower panels, respectively. The data are taken from
         Ref.~\cite{ndata,Roger}.  The circles correspond to the free result,
         while the squares and triangles include the medium contributions
         using the DH and HFB mean field potentials, respectively.
         The calculations below 350 MeV are based on
         the full Bonn NN t-matrix \cite{Bonn}, those above 350 MeV
         use the high energy extension of
         the Bonn potential \cite{D52} as input.  The asterisks
         represent $A\times\langle\sigma_t\rangle$, where
         $\langle\sigma_t\rangle$ is the isospin averaged neutron-nucleon
         total cross-section and $A$ is the atomic number,
         and the horizontal arrows give the geometric black disk cross
         section as described in the text. \label{fig2}}

\figure{ The isospin averaged neutron-nucleon total cross-section
         calculated from the average of the $(n,p)$ and $(n,n)$ results
         and shown as open circles
         is compared to the $^{16}$O and $^{40}$Ca neutron total
         cross-section measurements divided by $A_{eff}$.
         The solid line represents the $^{16}$O data divided by
  $A_{eff}=12$ and the dashed line the $^{40}$Ca data divided by
 $A_{eff}=26$.
         \label{fig3}}

\figure{ The elastic and reaction cross-sections are shown for $^{16}$O
         in the upper and lower panels, respectively.  The circles
         represent the free result, while the squares and triangles include
         the medium contributions using the DH and HFB mean field
         potentials, respectively.  \label{fig4}}

\figure{ The angular distribution of the differential cross-section
         ($\frac{d\sigma}{d\Omega}$), analyzing power ($A_y$) and
         spin rotation function ($Q$) for elastic neutron scattering
         from $^{16}$O at 100 MeV laboratory energy.  The
         calculations are performed with a first-order optical potential
         obtained from the full Bonn interaction \cite{Bonn}
         in the optimum factorized form.  The solid curve represents
         the free impulse approximation with using the free NN t-matrix.
         The medium contributions are included in the dashed curves,
         where the DH mean potential is used for the dashed-dotted  curve
         and the HFB mean field potential for the dashed
         curve. \label{fig5}}

\noindent
\figure{ Same as Fig.~\ref{fig5}, except the projectile kinetic energy
         is 500~MeV.  \label{fig6}}


\begin{references}

\bibitem{medium} C.R.~Chinn, Ch.~Elster, and R.M.~Thaler,
Phys. Rev. {\bf C48}  2956 (1993).

\bibitem{PT} A. Picklesimer and R. M. Thaler, Phys. Rev. {\bf C 23},
            42 (1981).
\bibitem{Sicil} E. R. Siciliano and R. M. Thaler, Phys. Rev. {\bf C 16},
                1322 (1977).

\bibitem{Bonn} R. Machleidt, K. Holinde, and Ch. Elster, Phys. Rep.
{\bf 149}, (1987).
\bibitem{D52} Ch. Elster and P.C. Tandy, Phys. Rev. {\bf C40} (1989),
881; Ch. Elster, Ph.D. thesis, University of Bonn, 1986.

\bibitem{HFB}See for example J.F. Berger, M.~Girod, and D.~Gogny,
 Nucl. Phys. {\bf A502}, 85c (1989); J.P.~Delaroche, M.~Girod, J.~Libert
 and I.~Deloncle, Phys. Lett. {\bf B232}, 145 (1989).
\bibitem{Gogny} J.F.~Berger, M.~Girod, and D.~Gogny, Comput. Phys.
    Commun. {\bf 63}, 365 (1991).

\bibitem{DH} Horowitz and B. Serot, Nucl. Phys {\bf A368}, 503 (1981).

\bibitem{Wolfe}A.~Picklesimer, P.C.~Tandy, R.M.~Thaler and D.H.~Wolfe,
Phys. Rev. {\bf C30}, 2225 (1984).

\bibitem{Ullman} P.W.~Lisowski et {\it al.}, Phys. Rev. Lett.
{\bf 49}, 255 (1982).

\bibitem{ndata}  R. W. Finlay, W. P. Abfalterer, G.~Fink, E.~Montei,
                 T~Adami, P.~W.~Lisowski, G.~L.~Morgan and R.~C.~Haight,
                 Phys. Rev.  {\bf C 47}, 237 (1993).

\bibitem{Roger} R.W.~Finlay, G.~Fink, W.~Abfalterer, P.~Lisowski,
G.L.~Morgan, and R.C.~Haight, in {\it Proceedings of the Internat.
Conference on Nuclear Data for
Science and Technology}, edited by S.M.~Qaim (Springer-Verlag, Berlin,
1992), p. 702.


\end{references}
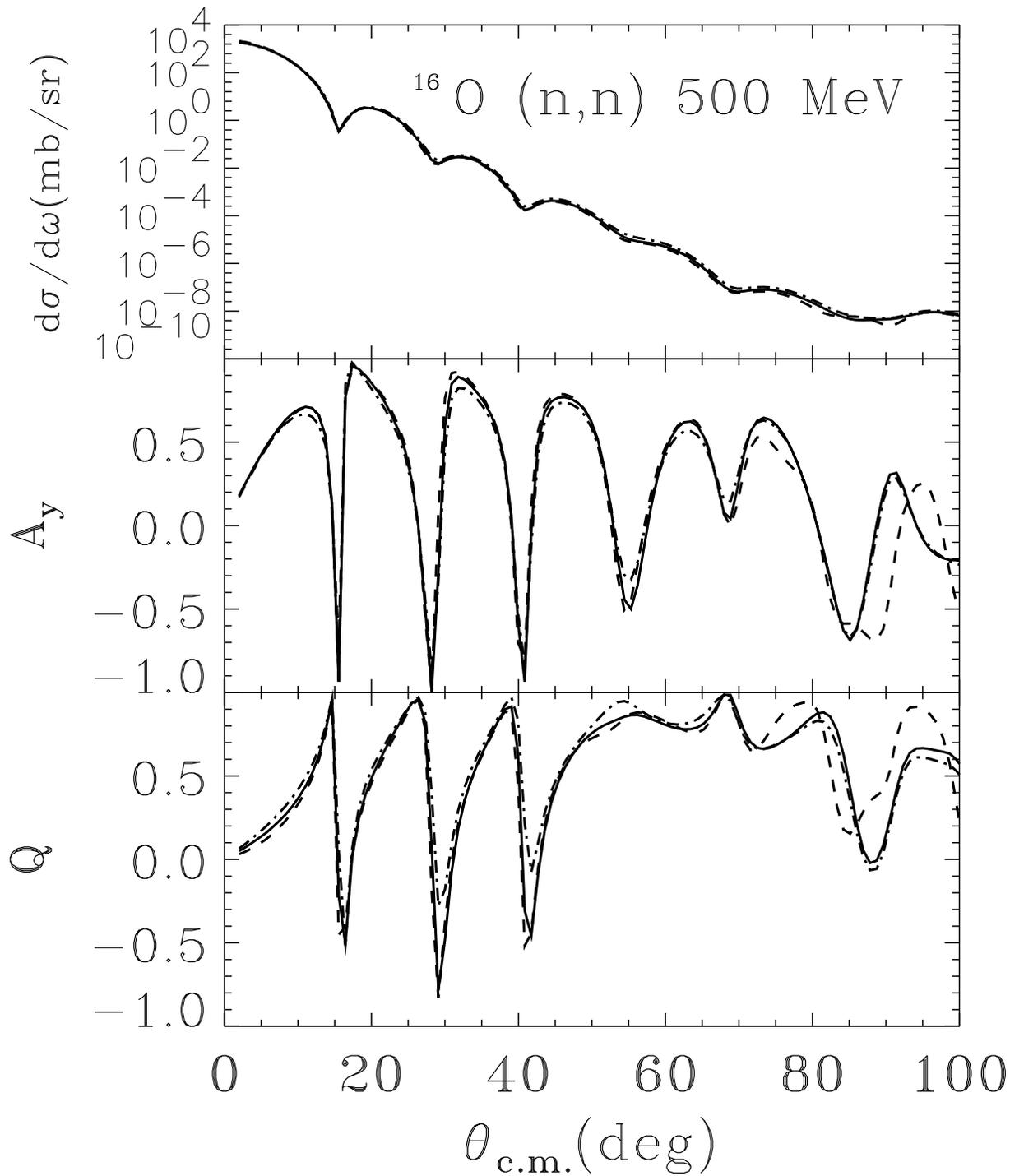
\end{document}